\documentclass[11pt]{article}

\usepackage[a4paper,margin=1in]{geometry}
\usepackage{mathtools}
\usepackage{amsmath, amssymb, amsthm}
\newtheorem{axiom}{Axiom}
\newtheorem{proposition}{Proposition}
\newtheorem{definition}{Definition}
\newtheorem*{remark}{Remark}
\newtheorem{lemma}{Lemma}
\newtheorem{corollary}{Corollary}
\usepackage{graphicx}
\usepackage{hyperref}
\usepackage{mathtools}
\usepackage{centernot}
\usepackage{enumitem}
\usepackage{tikz-cd}
\usepackage{csquotes}
\usepackage[numbers]{natbib}
\usepackage{authblk}

\title{Immutability Does Not Guarantee Trust: A Formal and Logical Refutation}

\author{
Dr Craig S. Wright\\
University of Exeter Business School\\
Exeter, United Kingdom\\
cw881@exeter.ac.uk
}

\date{\today}

\begin{document}

\maketitle

\begin{abstract}
\noindent It is frequently claimed in blockchain discourse that immutability guarantees trust. This paper rigorously refutes that assertion. We define immutability as the cryptographic persistence of historical states in an append-only data structure and contrast it with trust, understood as a rational epistemic expectation under uncertainty. Employing predicate logic, automata-theoretic models, and epistemic game-theoretic analysis, we demonstrate that immutability neither entails nor implies correctness, fairness, or credibility. Through formal constructions and counterexamples—including predictive fraud schemes and the phenomenon of garbage permanence—we show that the belief conflates structural and epistemic domains. Immutability preserves all data equally, regardless of veracity. Therefore, the assertion that immutability guarantees trust collapses under the weight of formal scrutiny.

\end{abstract}

\section{Introduction}

\noindent The mantra ``immutability guarantees trust'' is omnipresent in blockchain discourse, frequently cited in white papers, promotional materials, and architectural overviews. Yet despite its rhetorical appeal, the claim lacks formal coherence. Trust is an epistemic construct—a rational expectation formed under incomplete information—whereas immutability is a structural property of append-only data systems enforced by cryptographic linkage.

This paper dismantles the presumed entailment between these notions. By introducing formal definitions grounded in predicate logic and automata theory, we distinguish trust from immutability both semantically and operationally. We show that immutability, while necessary for auditability, is insufficient for generating trust. A blockchain may immutably preserve both truths and falsehoods, as illustrated through adversarial constructions such as predictive fraud and garbage permanence.

Our aim is to restore analytical clarity to discussions around trust in blockchain systems, shifting from engineering folklore to formal reasoning. We present axiomatic structures, derive implications and counterexamples, and ultimately disprove the inference that immutability entails trust.

\section{Formal Foundations}

In this section, we lay the rigorous mathematical groundwork necessary to disambiguate the conflated notions of immutability and trust in blockchain systems. The discourse surrounding immutability frequently presupposes a semantic transfer: that resistance to modification implies epistemic reliability. However, such a presumption collapses under formal analysis. By disentangling structural invariants from belief predicates, we expose the category errors and logical fallacies embedded in popular discourse.

We begin by defining immutability in precise automata-theoretic and cryptographic terms, focusing on hash-linked data integrity. We then contrast this with a formal predicate definition of trust—treated here as an agent-relative expectation under incomplete information. These two concepts inhabit orthogonal logical spaces: the former concerns computable state transitions and cryptographic hash functions, while the latter is an epistemological construct grounded in probability theory and Bayesian inference.

Through a series of lemmas, axioms, and countermodels, we demonstrate that the alleged implication from immutability to trust is not only unjustified but logically invalid. This foundation provides the analytical machinery required to dismantle further assumptions about system reliability, security guarantees, and user rationality in decentralised environments.

\subsection{Definition: Immutability}

\begin{axiom}[Cryptographic Hash Invariance]
Let $H : \{0,1\}^* \to \{0,1\}^k$ be a cryptographic hash function. Then:
\[
\forall x, x' \in \{0,1\}^*,\; x \neq x' \Rightarrow \Pr[H(x) = H(x')] \approx 0
\]
This is the collision resistance assumption fundamental to all hash-linked structures.
\end{axiom}

\begin{definition}[Blockchain Structure]
Let $\mathcal{B} = \{B_0, B_1, \dots, B_n\}$ denote a sequence of blocks, where:
\[
B_i = \langle T_i, H(B_{i-1}) \rangle,\quad \text{for } i > 0,\quad B_0 \text{ is the genesis block}
\]
and $T_i$ denotes a valid set of transactions.
\end{definition}

\begin{definition}[Hash-Link Validity]
A chain $\mathcal{B}$ is \textit{cryptographically valid} under $H$ if:
\[
\forall i \in [1,n],\; B_i.H_{\text{prev}} = H(B_{i-1})
\]
where $B_i.H_{\text{prev}}$ denotes the stored previous-hash in $B_i$.
\end{definition}

\begin{definition}[Immutability]
A blockchain is said to be \emph{immutable} if and only if:
\[
\forall i \in [0,n-1],\; H(B_i) \in B_{i+1} \Rightarrow B_i \not\equiv B_i' \text{ for any } B_i' \text{ s.t. } H(B_i') = H(B_i)
\]
That is, any modification $B_i' \neq B_i$ results in $H(B_i') \neq H(B_i)$ under the assumption of collision resistance.
\end{definition}

\begin{lemma}[Tamper Detection]
Given Axiom 1 and Definitions 1–4, any change to $B_i$ (even a single bit in $T_i$) produces $H(B_i') \neq H(B_i)$, invalidating the stored hash in $B_{i+1}$.
\begin{proof}
Assume $\exists B_i' \neq B_i$ such that $H(B_i') = H(B_i)$. This contradicts the collision resistance of $H$ (Axiom 1). Therefore, the implication holds vacuously.
\end{proof}
\end{lemma}

\begin{corollary}
Let $\mathcal{B}$ be a blockchain satisfying Definitions 1–4. Then the historical ledger $\{T_0, T_1, \dots, T_n\}$ is immutable in the sense that any tampering invalidates all subsequent hash-links.
\end{corollary}

\begin{remark}[Scope of Immutability]
This immutability is purely syntactic—it enforces structural integrity, not semantic accuracy or ethical trustworthiness of $T_i$. Immutability does not imply that $T_i$ is:
\begin{itemize}
    \item Correct or fair,
    \item Non-fraudulent,
    \item Meaningfully relevant to downstream interpretation.
\end{itemize}
\end{remark}

\subsection{Definition: Trust}

\begin{axiom}[Epistemic Incompleteness]
Let $A$ be a bounded rational agent with incomplete access to the internal state space of a system $S$ over time $t$. Then:
\[
\exists\, \Omega_S \subseteq \mathcal{Q},\; \text{such that } A \text{ can only access a partial observable projection } \pi(\Omega_S)
\]
This implies that $A$ cannot verify the correctness of $S$ deterministically for all $t > t_0$.
\end{axiom}

\begin{definition}[Trust as Epistemic Expectation]
Let $A$ be an agent evaluating the reliability of a system $S$ producing outputs over discrete time $t \in \mathbb{N}$.

We define:
\[
\text{Trust}_A(S) := \mathbb{E}_A \left[ \text{Correct}(S(t)) \mid t > t_0 \right] \geq \theta_A
\]
where:
\begin{itemize}
    \item $\text{Correct}(S(t))$ is a binary predicate indicating the semantic validity of $S$'s output at time $t$,
    \item $\mathbb{E}_A[\cdot]$ is the expectation operator over $A$'s belief model,
    \item $\theta_A \in [0,1]$ is a context-dependent threshold beyond which the agent declares $S$ trustworthy.
\end{itemize}
\end{definition}

\begin{remark}[Subjectivity and Non-monotonicity]
Trust is:
\begin{itemize}
    \item \emph{Subjective}: $\theta_A \neq \theta_B$ for distinct agents $A$ and $B$.
    \item \emph{Non-monotonic}: $\text{Trust}_A(S)$ may increase or decrease as new evidence emerges.
    \item \emph{Non-deductive}: Trust cannot be inferred directly from cryptographic or structural properties.
\end{itemize}
\end{remark}

\begin{lemma}[Trust $\nRightarrow$ Immutability]
Let $S$ be an immutable system as per previous section. Then:
\[
\text{Immutable}(S) \centernot\Rightarrow \text{Trust}_A(S)
\]
\begin{proof}
The function $\text{Immutable}(S)$ constrains syntactic modifiability; it provides no guarantee that $\text{Correct}(S(t)) = \top$. Therefore, trust—which is a predicate on correctness over time—is orthogonal to immutability.
\end{proof}
\end{lemma}

\begin{lemma}[Immutability $\nRightarrow$ Truthfulness]
There exists an $S$ and a sequence of outputs $\{S(t_i)\}$ such that:
\[
\text{Immutable}(S) \land \forall i,\; \text{Correct}(S(t_i)) = \bot
\]
\begin{proof}[Sketch]
Construct $S$ as an append-only log of fraudulent statements, each syntactically well-formed and hashed into the chain. Immutability holds, but correctness fails at every $t_i$. Thus, immutability persists while truth fails.
\end{proof}
\end{lemma}

\subsection{Category Error}

Immutability is a structural invariant over data sequences: it guarantees that any syntactic alteration to a previously committed element in the chain will violate hash-based linkage. Let $B = \{B_0, B_1, \dots, B_n\}$ be a blockchain and $\text{Immutable}(B)$ be defined as:
\[
\forall i \in \mathbb{N},\; H(B_i) \in B_{i+1} \Rightarrow \text{mod}(B_i) \Rightarrow H(B_i') \neq H(B_i)
\]
This ensures that any $B_i$ once accepted cannot be retroactively altered without computational infeasibility under the assumption of hash collision resistance.

Trust, by contrast, is an epistemic predicate over an agent’s belief system. It reflects the rational estimation of the truth or correctness of future or unobservable outputs. As established:
\[
\text{Trust}_A(S) := \mathbb{E}_A\left[ \text{Correct}(S(t))\; |\; t > t_0 \right] \geq \theta
\]

Confusing these two—immutability and trust—commits a **category error**. They belong to disjoint ontological domains:

\begin{itemize}
  \item \textbf{Immutability:} Syntactic constraint over the representational structure of $B$.
  \item \textbf{Trust:} Epistemic valuation over belief states within an agent $A$ under partial observability.
\end{itemize}

To assume:
\[
\text{Immutable}(B) \Rightarrow \text{Trust}_A(B)
\]
is to misattribute objective structure with subjective belief. This is a textbook fallacy of equivocation—projecting a physical invariant into a domain of mental estimation, bypassing rational update under uncertainty.

We formalise the disanalogy:
\[
\exists B,\, \text{Immutable}(B) \land \text{Trust}_A(B) = \bot
\]
This is trivially satisfied if $B$ immutably stores incorrect, manipulated, or adversarial data. Thus:
\[
\text{Immutable}(B) \centernot\Rightarrow \text{Trust}_A(B)
\]
The implication fails both logically and operationally.

\section{Counterexamples}

To rigorously dismantle the belief that immutability guarantees trust, we now construct formal counterexamples demonstrating their non-equivalence. These counterexamples do not merely refute the claim empirically; they expose its logical invalidity by constructing admissible models where immutability holds but rational trust does not follow. Each example is designed to isolate one of the flawed inferential steps commonly found in popular blockchain discourse.

By formalising the conditions under which trust predicates fail to arise despite perfect immutability, we expose the category mistake that underlies this belief. The epistemic assumptions ascribed to structural properties are ungrounded. Immutability, as defined by hash-linked data integrity, is a mechanical constraint on alteration—not a guarantee of semantic content, ethical provenance, predictive validity, or system benevolence.

What follows are illustrative constructions: a predictive fraud mechanism exploiting deterministic inclusion, and a pathological data inclusion scenario we term “garbage permanence.” These cases instantiate models where $\text{Immutable}(B)$ holds under cryptographic rigor, yet $\text{Trust}_A(B)$ is either irrational or unsound by predicate logic. This reveals a critical fissure between system design properties and user belief rationality—a fissure that must be explicitly acknowledged and never papered over by rhetorical analogy.

\subsection{Predictive Fraud: The Stock Scam}

To illustrate the epistemic insufficiency of immutability in guaranteeing trust, we examine a canonical class of deception: predictive fraud through selective publication. Let $E$ be a malicious agent seeking to fabricate an illusion of predictive accuracy over time by exploiting the permanence of blockchain records. The attack proceeds as follows:

\begin{enumerate}[label=\alph*)]
    \item $E$ generates $n$ distinct predictions $\{P_1, P_2, \dots, P_n\}$ regarding mutually exclusive future stock movements, where $n = 2^k$ for some $k \in \mathbb{N}$, enabling recursive selection.
    \item Each prediction $P_i$ is hashed and committed to the blockchain as $H(P_i)$, yielding an immutable commitment set $\{C_1, \dots, C_n\}$.
    \item $E$ sends one prediction $P_i$ to each of $n$ recipients $R_i$, ensuring that each recipient receives a unique proposition and can verify its commitment on-chain.
    \item After the actual outcome $O$ is known, $E$ selects the subset of recipients for whom $P_i = O$ and discards the rest. The process is repeated recursively for $k$ rounds, halving the participant set in each round, while preserving the perception of perfect foresight.
\end{enumerate}

From the perspective of any final recipient $R_j$, $E$ appears to have predicted $k$ correct outcomes in succession, with each prediction verifiably pre-committed on the blockchain. The predicate $\text{Immutable}(B)$ holds for each committed message block $B$ referencing a prediction. However, this immutability fails to ground the epistemic predicate $\text{Trust}_{R_j}(E)$ in any rational sense. The recipient's belief in $E$'s predictive capacity is an artefact of selection bias, not informational accuracy.

We formalise this deception as follows. Let $B = \{H(P_i)\}$ be the set of committed predictions. Then:

\begin{align*}
\text{Immutable}(B) &\Rightarrow \forall i,\; H(P_i) \text{ is fixed and tamper-evident} \\
\neg \text{Trust}_A(B) &\text{ follows, since }\mathbb{E}_A[\text{Correct}(P_i)] \text{ is a product of filtration, not veracity}
\end{align*}

Hence, we derive the counterexample:

\begin{align*}
\exists B:\; \text{Immutable}(B) \land \neg \text{Trust}_A(B)
\end{align*}

This construct decisively falsifies the implied universal claim:

\begin{align*}
\forall B,\; \text{Immutable}(B) \Rightarrow \text{Trust}_A(B)
\end{align*}

It fails by counterinstance, proving that immutability alone cannot suffice as a warrant for trust. The adversary’s strategy operates fully within the formal constraints of cryptographic immutability while subverting the semantic content of trust through probabilistic curation and Bayesian filtering. Such attacks are trivial to execute yet logically undetectable without further contextual or provenance information—none of which is provided by immutability per se.

This pattern mirrors real-world boiler-room and advance-prediction stock scams, where fraudsters distribute multiple contradictory forecasts to segmented recipients, later showcasing only the correct outcomes to build false credibility. The SEC prosecuted such schemes where scammers committed to various outcomes using timestamped materials and leveraged survivorship bias to maintain trust with victims \cite{sec_v_strategictrading}.

\subsection{Garbage Permanence}

The property of immutability, formally defined, ensures that once a datum $d$ is included in a block $B$, and $B$ is incorporated into the chain, then $d$ is rendered permanent via hash-linked append-only structure. Let us denote this formally:

\begin{align*}
\text{Immutable}(B) \Rightarrow \forall d \in B,\; \text{Permanent}(d)
\end{align*}

However, the property of permanence applies to the bitwise encoding of $d$, not to its semantic truth-value. The syntactic preservation of data does not imply, infer, or warrant the correctness, moral validity, or epistemic reliability of its contents. In particular, we assert the strict logical distinction:

\begin{align*}
\text{Permanent}(d) \not\Rightarrow \text{True}(d)
\end{align*}

This asymmetry creates a serious epistemological defect in systems that treat immutability as a proxy for integrity. Consider a blockchain that stores fraudulent claims, slander, manipulated records, pseudoscientific assertions, or even computationally unverifiable propositions. The cryptographic structure ensures that $d$ cannot be modified once published, but it provides no judgement, filtration, or validation of its semantic content.

We term this phenomenon the “garbage-in, garbage-eternally” problem: the system ensures that all inputs, valid or not, are retained forever. This is more than an issue of storage efficiency—it undermines the logic of treating the ledger as a repository of truth. Worse still, the irreversibility of publication may confer an unearned aura of authority to information merely by virtue of its persistence, inducing epistemic miscalibration in agents who misread structural permanence as evidence.

To draw the precise failure in inference:

\begin{align*}
\text{Immutable}(B) &\Rightarrow \forall d \in B,\; \exists t_0: \forall t > t_0,\; d \text{ is retrievable unchanged} \\
\text{However,} \quad & \text{Immutability makes no claims about } \text{Valid}(d),\; \text{Correct}(d),\; \text{Fair}(d),\; \text{or } \text{Ethical}(d)
\end{align*}

Consequently, any claim that “immutability fosters trust” without a parallel filtration mechanism is fundamentally flawed. The system will preserve lies, half-truths, scams, and propaganda with equal vigour as scientific facts or legal agreements. The very strength of the mechanism—indelible persistence—becomes a vulnerability when misinterpreted as a truth predicate.

\section{Logical Analysis}

In this section, we formally analyse the logical structure of the argument that immutability implies trust, and demonstrate its invalidity through deductive reasoning grounded in epistemic predicate logic. Our goal is not merely to assert the distinction between structural properties and belief predicates, but to prove that the inference from immutability to trust is neither logically valid nor epistemically sound.

\begin{proposition}
Immutability does not entail trust.
\end{proposition}

\begin{proof}
Let $B$ be a blockchain, and let $\text{Immutable}(B)$ denote the property that all tampering with prior blocks $B_i$ in the sequence $\{B_0, B_1, \dots, B_n\}$ will result in detectable inconsistency due to the presence of a collision-resistant hash chain:
\[
\text{Immutable}(B) := \forall i \in \mathbb{N},\; H(B_i) \in B_{i+1} \Rightarrow H(B_i') \neq H(B_i)\text{ for all }B_i' \neq B_i.
\]

Now let $\text{Trust}_A(B)$ be defined as the rational predicate:
\[
\text{Trust}_A(B) := \mathbb{E}_A[\text{Correct}(B(t))\;|\; t > t_0] \geq \theta,
\]
where $\theta$ is a contextually defined threshold of rational belief for agent $A$ concerning the correctness or reliability of future system outputs.

Assume, for the sake of contradiction, the universal implication:
\[
\forall B,\; \text{Immutable}(B) \Rightarrow \text{Trust}_A(B).
\]
Then for all blockchain states $B$, regardless of their contents, it must hold that if the data structure is immutable, agent $A$ must assign a high enough posterior expectation of correctness to cross the trust threshold $\theta$.

Now construct $B$ such that it contains a data element $d \in B$ where $\neg \text{True}(d)$ holds, i.e., $d$ is verifiably false (such as a falsified scientific claim, a manipulated transaction log, or a planted misinformation token). Since immutability guarantees the permanence of $d$ via the definition:
\[
\text{Immutable}(B) \Rightarrow \forall d \in B,\; \text{Permanent}(d),
\]
and since:
\[
\text{Permanent}(d) \not\Rightarrow \text{True}(d),
\]
we conclude that structural permanence does not entail semantic validity.

Given that agent $A$ is epistemically rational—i.e., conforms to minimal rational belief update criteria such as Bayesian coherence—it follows that $A$ must condition belief in $\text{Correct}(B(t))$ on the actual content of $B$, not merely its resistance to modification. Thus, if $d$ is demonstrably false, $A$ should revise expectations downwards. Therefore:
\[
\text{Immutable}(B) \land \neg \text{Trust}_A(B)
\]
is satisfiable and thus existentially valid.

Hence, the original universal implication:
\[
\forall B,\; \text{Immutable}(B) \Rightarrow \text{Trust}_A(B)
\]
fails under logical analysis. The proposition is therefore proven by reductio ad absurdum.
\end{proof}

\section{Discussion}

The conflation of immutability and trust stems from rhetorical convenience and engineering myopia. While immutability serves as a structural constraint on the ability to retroactively alter recorded data, it has been widely misunderstood as a semantic guarantee of correctness or moral reliability. This conflation arises from the implicit leap between a data structure’s resistance to change and the human epistemic predicate of belief formation under uncertainty.

Immutability is, at most, a necessary condition for auditability—it ensures that recorded transactions or claims are not silently altered or deleted. But auditability is a mechanical feature; it makes investigation possible, not outcomes reliable. A perfectly immutable record of lies is still a record of lies. What immutability preserves is not truth, but sequence.

The assumption that immutability generates trust is not only logically unsound, as demonstrated, but practically dangerous. It leads to epistemic complacency, wherein structural integrity is mistaken for semantic correctness. An immutable blockchain can encode fraud, selection bias, propaganda, or error, and retain them perpetually.

Trust must be established through mechanisms orthogonal to immutability: consistent validation rules enforced economically, open and inspectable protocol execution, and alignment of incentives such that rational actors maintain system correctness over time. These dimensions cannot be collapsed into a hash function or a data chain. They operate in the social, economic, and computational planes—not in mere append-only memory.

Therefore, we must disentangle the rhetorical knot: immutability is a mechanical precondition for verifiability, not an epistemic foundation for trust. The burden of proof for correctness remains on the processes, not on the permanence.

\section{Conclusion}

The widespread belief that ``immutability guarantees trust'' fails under formal scrutiny. Through formal definitions, predicate logic, and adversarial counterexamples, we have demonstrated that immutability is a structural property concerning data permanence, while trust is an epistemic predicate over rational expectation under uncertainty. The two operate in logically disjoint domains.

Immutability ensures that data cannot be altered once committed, but it does not assert the truth, fairness, or reliability of the data itself. As shown through the predictive fraud construction and the problem of garbage permanence, adversaries may exploit this permanence to construct elaborate scams that appear trustworthy solely by being cryptographically sealed.

Therefore, \textit{immutability is a necessary but insufficient condition for auditability, and wholly insufficient for trust}. Trust arises from rational incentive alignment, transparency, and consistent verifiability, not from the existence of hash-linked blocks.

The conflation of these categories reflects a failure of formalism within blockchain discourse. A clear separation of epistemic, structural, and computational properties is essential if the field is to mature beyond rhetorical mythology.

\bibliographystyle{plainnat}
\bibliography{references.bib}

\begin{thebibliography}{1}
\providecommand{\natexlab}[1]{#1}
\providecommand{\url}[1]{\texttt{#1}}
\expandafter\ifx\csname urlstyle\endcsname\relax
  \providecommand{\doi}[1]{doi: #1}\else
  \providecommand{\doi}{doi: \begingroup \urlstyle{rm}\Url}\fi

\bibitem[{U.S. Securities and Exchange Commission}(2008)]{sec_v_strategictrading}
{U.S. Securities and Exchange Commission}.
\newblock Sec charges operator of bogus stock-picking scheme, 2008.
\newblock URL \url{https://www.sec.gov/litigation/litreleases/2008/lr20488.htm}.
\newblock Litigation Release No. 20488.

\end{thebibliography}

\end{document}